\documentclass{aa}
\usepackage{graphicx}
\usepackage{natbib}
\bibpunct{(}{)}{;}{a}{}{,}
\begin{document}
   \title{The influence of redshift information on galaxy-galaxy lensing
          measurements}

   \author{M. Kleinheinrich\inst{1,2}
          \and H.-W. Rix\inst{1}
          \and T. Erben\inst{2}
          \and P. Schneider\inst{2}
          \and C. Wolf\inst{3} 
          \and M. Schirmer\inst{2}
          \and K. Meisenheimer\inst{1}
          \and A. Borch\inst{1}
          \and S. Dye\inst{4}
          \and Z. Kovacs\inst{1}
          \and L. Wisotzki\inst{5}
          }

   \offprints{M. Kleinheinrich,\\ \email{martina@mpia.de}}

   \institute{Max-Planck-Institut f\"ur Astronomie, K\"onigstuhl 17, 
              D-69117 Heidelberg, Germany 
              \and 
              Institut f\"ur Astrophysik und Extraterrestrische Forschung, 
              Universit\"at Bonn, Auf dem H\"ugel 71, 53121 Bonn, Germany
              \and
              Department of Physics, Denys Wilkinson Bldg., University of 
              Oxford, Keble Road, Oxford, OX1 3RH, U.K.
              \and 
              School of Physics and Astronomy, Cardiff University, 5 The 
              Parade, Cardiff, CF24 3YB, U.K.
              \and
              Astrophysikalisches Institut Potsdam, An der Sternwarte 16, 
              D-14482 Potsdam, Germany
             }

   \date{Received / Accepted}

   \abstract{We investigate how galaxy-galaxy lensing measurements depend on
             the knowledge of redshifts for lens and source
             galaxies. Galaxy-galaxy lensing allows one to study dark matter
             halos of galaxies statistically using weak gravitational lensing.
             Redshift information is required to reliably distinguish
             foreground lens galaxies from background source galaxies and to
             convert the measured shear into constraints on the lens model. 
             Without spectroscopy or multi-colour information, redshifts can
             be drawn from independently estimated probability distributions.
             The COMBO-17 survey provides redshifts for both lens and
             source galaxies. It thus offers the unique possibility to do this
             investigation with observational data. We find that it is of
             great importance to know the redshifts of individual lens 
             galaxies in order to constrain the properties of their dark 
             matter halos. Whether the redshifts are derived from $UBVRI$ or
             the larger number of filters available in COMBO-17 is not very
             important. In contrast, knowledge of individual source
             redshifts improves the measurements only very little over the use
             of statistical source redshift distributions as long as the
             source redshift distribution is known accurately.

   \keywords{Gravitational lensing -- Methods: data analysis -- Galaxies: 
             fundamental parameters -- Galaxies: statistics -- Cosmology: dark 
             matter}
   }

   \maketitle
%

\section{Introduction}
\label{sect: introduction}

Weak gravitational lensing of galaxies by galaxies (galaxy-galaxy lensing)
provides a unique tool to study the dark matter distribution in lens
galaxies. A foreground galaxy distorts the images of background galaxies such
that they are, on average, tangentially aligned with respect to the lens. This
alignment can be measured statistically, see e.g.\
\citet{mellier1999,bartelmann2001} for reviews. 

In the beginning, galaxy-galaxy lensing was measured from imaging data alone
without any direct redshift information
\citep[e.g.][]{brainerd1996,dellantonio1996,fischer2000}. However,
multi-colour or spectroscopic data are desirable for two reasons:
\begin{enumerate}
  \item The strength of the gravitational shear and thus the image distortion
    depends on the angular diameter distances between observer, lens and
    source. In turn, redshift estimates for the lens and source galaxies are
    therefore needed to translate the measured distortion reliably into
    constraints on the lens galaxies.
  \item The need for averaging over at least hundreds of lens galaxies
    complicates the interpretation of the results severely unless physically
    similar subsets of galaxies can be identified. As the luminous parts of
    galaxies cover a large range of properties, it is clear that galaxies
    also differ in the properties of their dark matter halos
    \citep[e.g.][]{mckay2002,prada2003}. To learn something about different
    galaxy types one needs to be able to classify the potential lens galaxies
    at least by their luminosities and rest-frame colours, which requires
    redshift estimates. Note, that for this appliciation the redshift
    precision required is only a few percent, not the $10^{-4}$ of
    spectroscopic surveys. 
\end{enumerate}
Redshifts and/or classification for the lens galaxies have been used by 
several authors now
\citep[e.g.][]{smith2001,wilson2001,mckay2001,guzik2002,hoekstra2003}, while
redshifts for the source galaxies have so far only been used by
\citet{hudson1998} who measured galaxy-galaxy lensing in the Hubble Deep Field
North, and more recently by \citet{sheldon2004} using the Sloan Digital Sky
Survey (SDSS). All other studies used redshift probability distributions at a
given apparent magnitude to assign redshifts to the source and -- where
necessary -- to the lens galaxies. These probability distributions have been
derived from galaxy redshift surveys like the Canada-France Redshift Survey
\citep{crampton1995} or the Caltech Faint Galaxy Redshift Survey
\citep{cohen2000}. However, for the deeper data sets, the measured redshift
probability distributions had to be extrapolated to the fainter magnitudes of
the source galaxies. 

In this paper we address the question of how measurements of dark matter halos
with galaxy-galaxy lensing are affected by incomplete redshift information. We
use observational data with accurate photometric redshifts for both lens and
source galaxies. We perform the measurements with (a) the full information for
lens and source galaxies, (b) full information for the lens galaxies but
drawing the redshifts of the sources from probability distributions, 
and (c) using probability distributions to estimate redshifts for
lenses and sources. Although in case (a) we use the full redshift information,
this measurement is not an optimal measurement of galaxy-galaxy lensing from
COMBO-17. In this paper, the lens and source selection is always based
on apparent magnitudes and angular separations which allows us to
compare the results from the different cases. However, for retrieving
the tightest and most meaningful constraints on the dark matter halos
of galaxies, one should clearly use redshifts also for the lens and
source selection. This will be done in a companion paper
\citep{kleinheinrich2004}. 

In Sect.\ \ref{sect:data} we describe the data set followed by an
overview of our method of measuring galaxy-galaxy lensing in Sect.\
\ref{sect:method}. Section \ref{sect:z_estimates} explains how we estimate
redshifts and luminosities. In Sect.\ \ref{sect:measurements} we present
our measurements and results. We close with a summary in Sect.\
\ref{sect:outlook}.

\section{Data}
\label{sect:data}
We use the COMBO-17 survey for our lensing analysis. Properties of the galaxy
sample have already been described in \citet{wolf2003}. A more detailed and
updated description of the final catalogs used here is given in
\citet{wolf2004}. COMBO-17 has already been applied to various weak lensing 
studies including detailed analysis of the supercluster Abell 901/902
\citep{gray2002,gray2004,taylor2004} and cosmic shear
\citep{brown2003,bacon2004}.

COMBO-17 is a deep optical survey carried out with the Wide Field Imager (WFI)
at the MPG/ESO 2.2-m telescope on La Silla, Chile. It consists of four fields
covering about 0.26 square degrees each and has in its deepest stacked R-band
image a 5-$\sigma$ point source limit of $R\sim 26$. Observation in $UBVRI$
and 12 optical medium-band filters yield spectral classification and
photometric redshifts with $\sigma_z/(1+z)<0.01$ for $R<21$, and deteriorating
to $\sigma_z/(1+z)\approx 0.05$ for $R\approx24$. The $R$-band observations
were carried out in the best seeing conditions (typically $0\farcs 75$)
enabling accurate shape measurements for weak lensing studies. Currently, the
data set is fully processed for three survey fields \citep[see][]{wolf2003},
which will all be used here. Although in principle all data needed for our
analysis is available from the standard COMBO-17 data reduction, we only use
those data for measurements based on photometry. This includes apparent
magnitudes, redshift estimates, classification and the derivation of
rest-frame luminosities. The data reduction pipeline of COMBO-17 is optimized
for photometric measurements, but it creates only simple sum images for object
detection and crude shape measurements. In particular, the sum images in
COMBO-17 were created using only full pixel shifts when stacking individual
exposures. Therefore, they are not optimal for shape measurements. Instead, we
detect objects and measure their shapes from coadded $R$-band images created
with the data reduction pipeline of the Garching-Bonn Deep Survey
\citep[GaBoDS,][]{schirmer2003}. In the coaddition of the image on the CDFS
field we included observations from the ESO Imaging Survey and from GOODS. The
exposure times and the seeing are 57000~s and $0\farcs88$ for the CDFS field,
21600~s and $0\farcs88$ for the S 11 field, and 24900~s and $0\farcs74$ for
the A 901 field. From these summed images we measure positions, shapes and
half-light radii. 

Note that the summed images used here are not identical to those used
in previous COMBO-17 lensing studies listed above. Also, application
of the KSB algorithm for shape measurement differs slightly (see
Sect.\ \ref{sect:shapes}). A joint effort to compare the results from
the different analyses is currently underway (Heymans et al.\ in
prep.).

The original field selection in COMBO-17 was not random. The A 901
field was chosen because of the presence of a supercluster composed of
Abell 901a,b and Abell 902 at $z=0.16$. Conversely, the CDFS field was
originally chosen because of its emptiness. The S 11 field is the
only random field. In \citet{kleinheinrich2004} we will present a
detailed analysis showing that the clusters in the A 901 field do
not affect the galaxy-galaxy lensing measurement. Therefore, in this
paper we completely ignore the presence of these clusters.

\subsection{Shape measurements}
\label{sect:shapes}
The coaddition of individual images into deep summed frames with the GaBoDS
pipeline (\textit{THELI}) is described in detail in
\citet{schirmer2003}. This pipeline was specifically developed for
data from multichip cameras such as the WFI with $2\times 4$
chips. The stacked images cover the whole CCD mosaic unlike having
stacked images for each individual chip. However, due to e.g.\
different sensitivities of the CCDs and gaps between them, the
exposure times and noise properties vary across the stacked
images. Therefore, an additional weight image is created that keeps
track of these varying noise properties. The object catalog is
obtained by running SExtractor \citep{bertin1996} on the stacked
images using the weight images. The use of the weight images very
efficiently suppresses spurious source detections. We therefore do not
apply further masking of objects by hand.

For the detected objects we use the KSB algorithm
\citep{kaiser1995,luppino1997,hoekstra1998} to measure shapes and correct
these for the point spread function (PSF). For such a correction we use the
shapes of stellar objects to trace the shape and variation of the PSF across
the image; see also \citet{bartelmann2001} for a summarizing description of
the KSB algorithm and \citet{erben2001} for details of its application as used
here.

Ellipticities are calculated from weighted second order-moments $Q_{ij}$ of
the light distribution $I(x)$  
\begin{equation}
\label{eq:Qij}
   Q_{ij}=\int d^2 x W(x)x_{i}x_{j}I(x)~.
\end{equation}
Here, $x$ denotes the position (a vector in the complex plane) with respect to
the object center, $W(x)$ is a weight function and $f(x)$ the flux at position
$x$. For the weight function $W(x)$ we use a gaussian filter function with the
half-light radius $\theta_h$ of the object under consideration (determined by
SExtractor) as the window scale. A complex ellipticity $\chi=\chi_{1}+i\chi_{2}$
is defined as  
\begin{equation}
\label{eq:chi}
   \chi_{1}=\frac{Q_{11}-Q_{22}}{Q_{11}+Q_{22}}, \quad
   \chi_{2}=\frac{2Q_{12}}{Q_{11}+Q_{22}} ~.
\end{equation}

The correction, $q$, for the spatially varying anisotropic part of the
PSF is applied by fitting a fourth-order polynomial to the
ellipticities of stellar objects across the whole field-of-view.

The subsequent correction for the isotropic part of the PSF is done as
a function of galaxy size. This means that the shear polarizability
$P^\mathrm{sh}$ and the smear polarizability $P^\mathrm{sm}$ of stellar
objects are measured with a range of window scales. For each object, the
measurement from the window scale matching $\theta_h$ of that object is used
in the correction. $P^\mathrm{sh}$ and $P^\mathrm{sm}$ are tensors that
describe, for a given weight function $W$, the response of the measured
ellipticities to gravitational shear in the absence of PSF effects
($P^\mathrm{sh}$) and to the isoptropic smearing by the PSF
($P^\mathrm{sm}$). $P^\mathrm{sh}$ and $P^\mathrm{sm}$ are measured from the
third and fourth-order moments of the light distribution of an object. An
estimator of the shear is then given by  
\begin{equation}
\label{eq:g}
  \epsilon=(P^{\mathrm{g}})^{-1}\chi_{\mathrm{aniso}}
\end{equation}
with 
\begin{equation}
\label{eq:Pg}
  P^{\mathrm{g}}=P^{\mathrm{sh}}-P^{\mathrm{sm}}(P^{*\mathrm{sm}})^{-1}P^{*\mathrm{sh}}~.
\end{equation}
The quantities with asterisks are measured from the stellar objects
only, and $\chi_\mathrm{aniso}=\chi+P^\mathrm{sm}q^*$ is the
anisotropy-corrected ellipticity. Instead of using full tensors in
Eqs.\ (\ref{eq:g}) and (\ref{eq:Pg}) we apply the scalar correction by
only using the trace of $P^\mathrm{g}$. Furthermore, we use raw values
of $P^\mathrm{g}$ although this leads in some cases to unphysically
large PSF-corrected ellipticities ($\epsilon>1$) due to
noise. However, \citet{erben2001} found that fitting $P^\mathrm{g}$
does not improve the shape measurement. We use the weighting scheme
proposed by \citet{erben2001} to downweight objects with noisy
ellipticity estimates. This procedure will be explained in more detail
in Sect.\ \ref{sect:method}.

\section{Method}
\label{sect:method}
For extracting the galaxy-galaxy lensing signal from a set of measured galaxy
ellipticities and, potentially, redshifts, we use the maximum-likelihood
method by \cite{schneider1997}. We parametrize the lens galaxies as singular
isothermal spheres
\begin{equation}
\label{eq:SIS_rho}
   \rho(r)=\frac{\sigma_v^{2}}{2\pi G} \frac{1}{r^2} ~,
\end{equation}
where $\sigma_v$ is the velocity dispersion of a galaxy. 
Furthermore,  we assume that the velocity dispersion scales with 
luminosity as
\begin{equation}
\label{eq:SIS_tf}
    \frac{\sigma_v}{\sigma_*}=\left(\frac{L}{L_*}\right)^\eta~,
\end{equation}
where $\sigma_*$ is the velocity dispersion of an $L_*$ galaxy.
We adopt $L_*=10^{10}L_{\sun}$, measured in the SDSS $r$-band.
For each lens-source pair the shear is given by
\begin{equation}
\label{eq:SIS_gamma}
    \gamma(r)=\frac{2\pi\sigma_{v}^{2}}{c^{2}}\frac{D_{\mathrm{ds}}}{D_{\mathrm{s}}}\frac{1}{\theta}~,
\end{equation}
where $D_{\mathrm{s}}$ is the angular diameter distance between the
observer and
source, and $D_{\mathrm{ds}}$  is that between the lens and source, while
$\theta$ is the angular separation between lens and source. For the
calculation of angular diameter distances from the galaxy redshifts,
we adopt $(\Omega_m,\Omega_\Lambda)=(0.3,0.7)$ and 
$H_0=100h~\mathrm{km~s}^{-1}\mathrm{Mpc}^{-1}$.

Following \citet{schneider1997}, for each source galaxy $j$ the total shear
$\gamma_j$ from all lenses lying within a certain annulus around the source is
calculated for a range of trial input parameters $(\sigma_*,\eta)$. The
resolution of our grid in parameter space is $\Delta
\sigma_\star=6~\mathrm{km/s}$ and $\Delta \eta=0.03$. From $\gamma_j$ and the
observed ellipticity $\epsilon_j$ the intrinsic ellipticity
\begin{equation}
\epsilon_{j}^{(s)}=\epsilon_{j}-\gamma_{j}
\label{eq:epsilon_intr}
\end{equation}
is estimated. The probability for observing this intrinsic ellipticity is
given by
\begin{equation}
P(\epsilon_{j}^{(s)})=\frac{1}{\pi\sigma_{\epsilon}^{2}} \exp{\left[-\frac{|\epsilon_{j}^{(s)}|^{2}}{\sigma_{\epsilon}^{2}}\right]}~,
\label{eq:P}
\end{equation}
where $\sigma_\epsilon$ is the width of the intrinsic ellipticity
distribution. Multiplying the probabilities $P(\epsilon_{j}^{(s)})$ from all
sources gives the likelihood for a given set of $(\sigma_*,\eta)$ which is
then maximized to find the best-fit parameters. 

The value of $\sigma_\epsilon$ can be estimated from the data set
itself. However, we find that the average ellipticity depends
systematically on, e.g., the signal-to-noise S/N or on the half-light
radius $\theta_h$ of the objects. This behaviour was suggested by the
weighting scheme proposed by \citet{erben2001} who assumed that the
noise properties of objects are traced by these quantities. They then
suggest the inverse of the variance $\sigma_\epsilon^2=\frac{1}{N}\sum
|\epsilon|^2$ as the weight of an object $j$ in the ellipticity
estimates of the $N\approx 20$ objects being closest to object $j$ in
$\theta_h-\mathrm{S/N}$ space. We use $\sigma_\epsilon$ from this
weighting scheme with $N=20$ as an estimate of the width of the
intrinsic ellipticity distribution for each object $j$. This choice of
$\sigma_\epsilon$ directly accounts for errors in the shape
measurement but also for intrinsic changes of $\sigma_\epsilon$ with
half-light radius and S/N which might be due to changes in the galaxy
population. Typical values are $\sigma_\epsilon\approx 0.4$.

The method of \citet{schneider1997} described so far only works if the
lens and source galaxy redshifts are known. If that is not the case,
one can calculate the shear $\gamma$ by integrating over a given
redshift probability distribution which must be done numerically. We
use Monte Carlo integration and assign random redshifts to the
galaxies 50 times, see
\citet{schneider1997}. The redshift probability distribution used is
described in Sect.\ \ref{sect:z_estimates}.

\section{Redshift distribution}
\label{sect:z_estimates}
For many of the COMBO-17 galaxies we have fairly precise redshifts. However,
for the majority of sources, with $R>23.5$, and for other data sets, we have
not. Therefore, we briefly describe how to create Monte Carlo redshifts, given
an apparent galaxy magnitude.

A parametrization of a redshift probability distribution which has often been
used in weak lensing is given by \citep{baugh1993} 
\begin{equation}
\frac{\mathrm{d}N}{\mathrm{d}z}\propto\frac{z^2}{z_\star^3}\exp\left[ -\left(\frac{z}{z_\star}\right)^{1.5}\right]~.
\label{eq:Prob_z}
\end{equation}
$z_\star$ is related to the median redshift $z_m$ of the distribution by
\begin{equation}
z_\star=z_m/1.412~.
\label{eq:z_median}
\end{equation}
\citet{brown2003} calculated median redshifts $z_m$ for different bins in
magnitude and found that in COMBO-17 $z_m$ increases with median $R$-band
magnitude as 
\begin{equation}
z_m=2.53162-0.329974\times R+0.0108296\times R^2~.
\label{eq:zm_mr}
\end{equation}
For each object we assign a random redshift by first calculating $z_m$ from
its $R$-band magnitude and using this $z_m$ in Eqs.\ (\ref{eq:Prob_z}) and
(\ref{eq:z_median}).

When we estimate the redshifts of the lens galaxies we also have to estimate
their luminosities. We use
\begin{equation}
L=\left( \frac{H_0 D_\mathrm{d}}{c} \right) ^2 (1+z)^6 10^{0.4(21.4-R)}
\label{eq:L}
\end{equation}
to relate the luminosity of a galaxy in units of $L_*$ to its apparent
$R$-band magnitude and redshift $z$ \citep{brainerd1996}. The exponent of the
$1+z$ term includes an estimate of the $k$-correction. We compared the
luminosities calculated from our measured redshifts and Equation (\ref{eq:L})
to those derived directly in COMBO-17, in order to test the accuracy of
Equation (\ref{eq:L}) and to determine its normalization. For 62\% of the
galaxies the luminosity estimate differs by less than 20\% from that measured
within COMBO-17.

\section{Measurements and results}
\label{sect:measurements}
For all measurements we constrain lenses to have magnitudes $R=18-22$ and
sources to $R=22.5-24$. Additionally, we require sources to have a
half-light radius larger than the PSF of the corresponding summed frame. The
maximum projected distance at which we still consider a lens acting on a 
source is $35\arcsec$, corresponding to an impact parameter of about
$130h^{-1}~\mathrm{kpc}$ for a typical lens redshift $z=0.4$. We find that
inclusion of pairs with larger projected distances reduces the significance of
our measurements. Furthermore, we exclude all pairs which are closer than
$8\arcsec$ because there the shape measurement of the source may be
affected by the light from the lens. Note that the average half-light radius
of the lens population is $0\farcs 85$ with a maximum of about
$3.5\arcsec$. Only objects identified as galaxies in the multi-colour
classification are used as lenses. Therefore, unresolved galaxies are
included in the lens sample.

Three main cases differing in sample size and extent of redshift information
are used to investigate the influence of redshift information on the
likelihood contours for the parameters $(\sigma_*,\eta)$: 

(a) Lens and source candidates must be classified as galaxies and have a
redshift measurement; redshifts from COMBO-17 are used for lenses and
sources.

(b) The lens sample is the same as in case (a) but sources can be all resolved
objects within the required magnitude range. Redshifts from COMBO-17 are used
for lenses whereas the redshifts of sources are drawn from probability
distributions based on their magnitudes.

(c) Lens and source sample are the same as in case (b) but redshifts from
COMBO-17 are not used at all. Lenses and sources are assigned random
redshifts based on their apparent magnitudes and using Eqs.\
(\ref{eq:Prob_z})--(\ref{eq:zm_mr}). Then, the luminosities of the lenses are
estimated from apparent magnitudes using Eq.\ (\ref{eq:L}).

In all cases we use the same lens sample. Therefore, any differences in the
parameter estimates must be due to either the accuracy of the redshifts or to
a change in the source sample. Note that we cannot straightforwardly use the
same source sample in all three cases as we do with the lens sample. This is
because the redshift probability distribution described in Sect.\
\ref{sect:z_estimates} was derived taking the completeness of the COMBO-17
redshifts into account, see also Sect.\ \ref{sect:sources_sub1}. 

\begin{figure*}
  \begin{center}
   \leavevmode
   \begin{minipage}[l]{0.32\textwidth}  
      \includegraphics[angle=-90,width=\hsize]{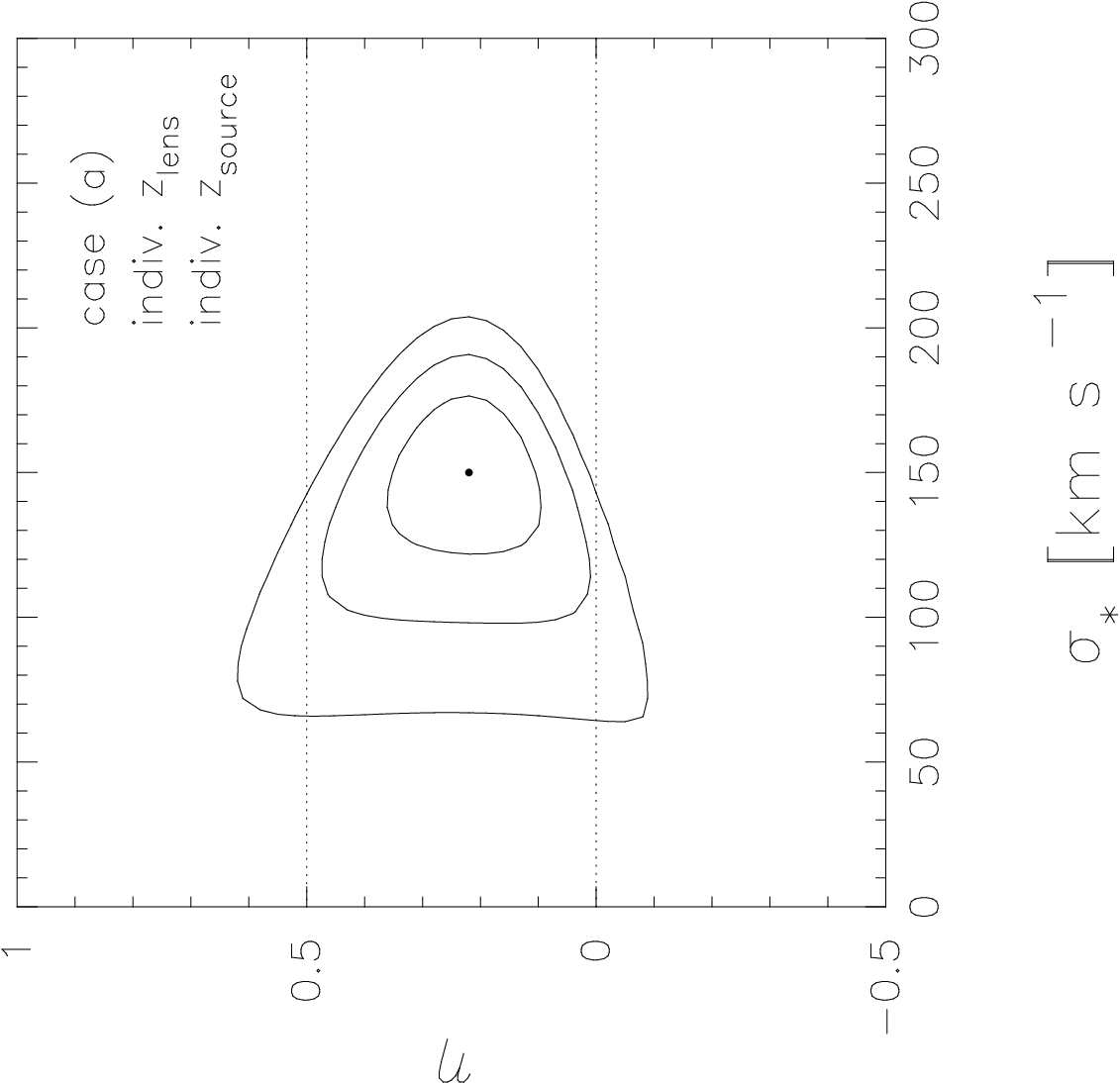}
   \end{minipage}
   \begin{minipage}[c]{0.32\textwidth}  
      \includegraphics[angle=-90,width=\hsize]{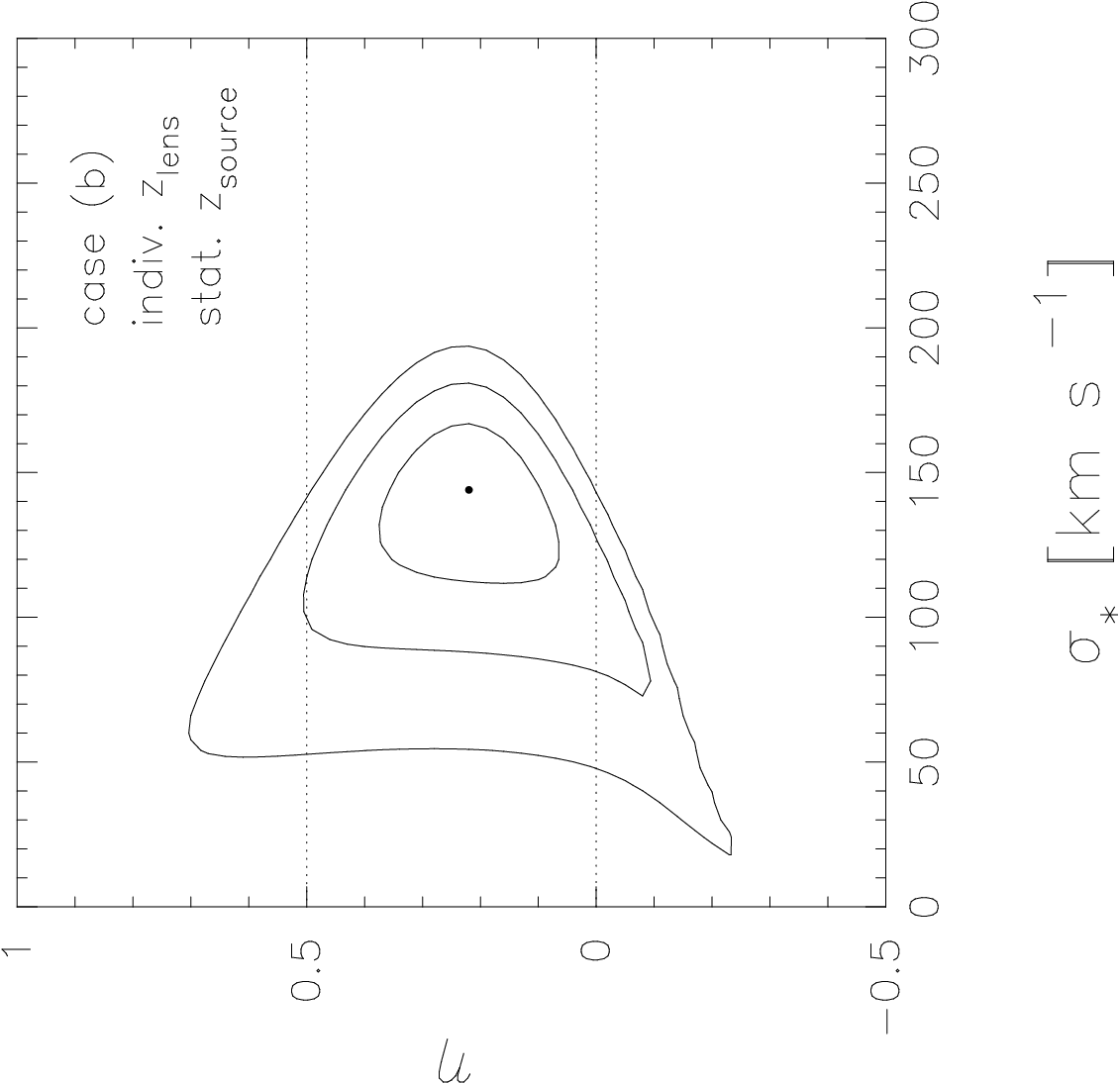}
   \end{minipage}
   \begin{minipage}[r]{0.32\textwidth}
      \includegraphics[angle=-90,width=\hsize]{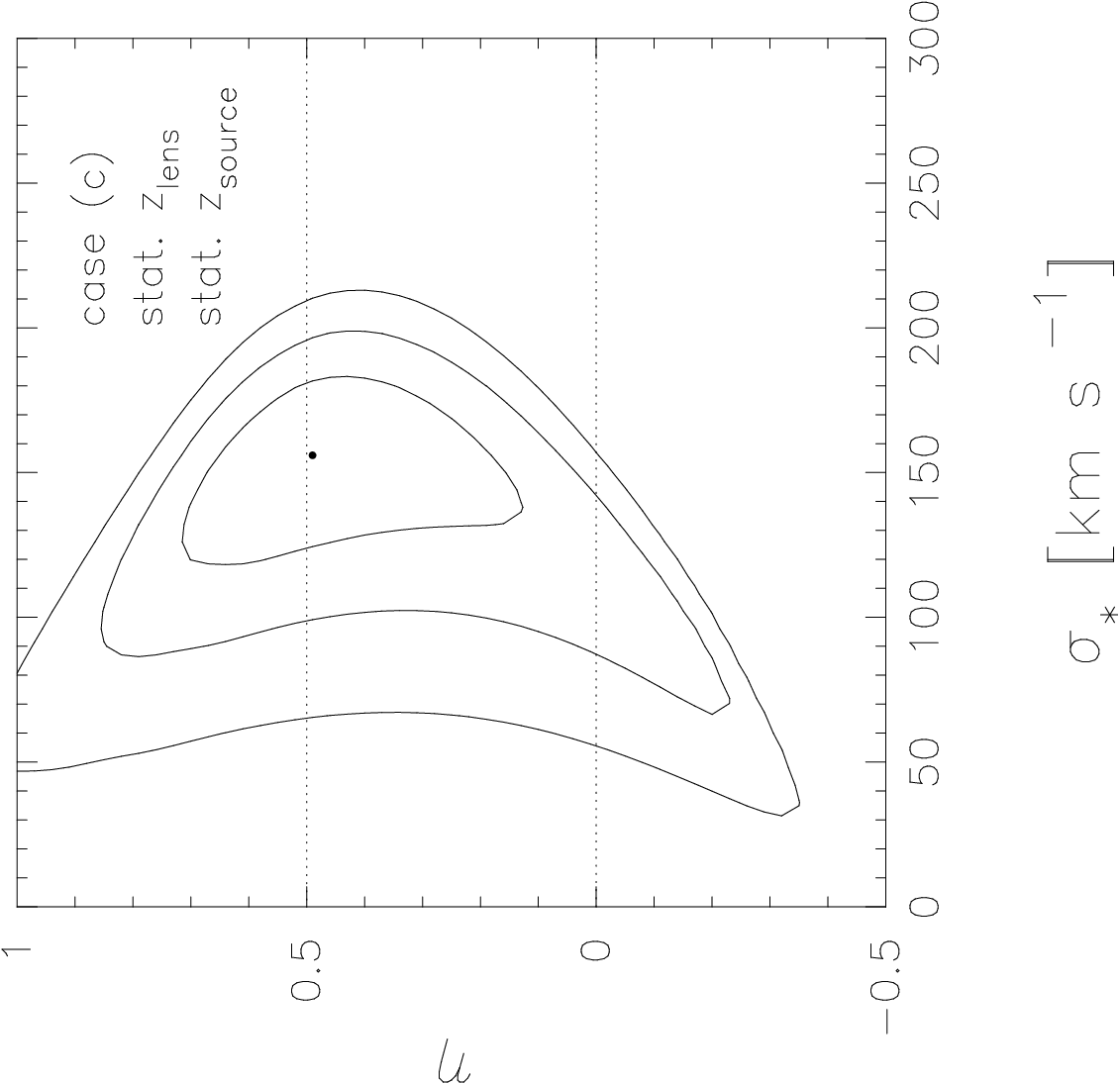}
   \end{minipage}

   \caption{Likelihood contours for the lens model described in Sect.\
     \ref{sect:method}, contours are 1-, 2- and 3-$\sigma$. Left panel:
     redshift information is used for lens and source galaxies, middle panel:
     redshift information is only used for lenses, right panel: redshift
     information is not used at all}
   \label{fig:measurements}
   \end{center}
\end{figure*}

Table \ref{tab:results} gives the number of lenses, sources and pairs used in
each measurement and the best-fit $(\sigma_*,\eta)$ values with their
1-$\sigma$ uncertainties. The corresponding likelihood contours are shown in
Fig.\ 
\ref{fig:measurements}.  

\begin{table}[htbp]
\begin{center}
\caption{Numbers of lenses $N_\mathrm{d}$, sources  $N_\mathrm{s}$ and pairs
         $N_\mathrm{p}$ used in the different measurements. For (b) and (c)
         $N_\mathrm{p}$ is only the number of pairs averaged over the
         different MC realizations, while $N_\mathrm{d}$ and $N_\mathrm{s}$ 
         give the number of lenses and sources that are used in any of the
         MC realizations. The last two columns give the best 
         fit values of $\sigma_*$ and $\eta$ with 1-$\sigma$ error bars
         marginalized over the other variable.}  
\begin{tabular}{c|r|r|r|r|r}
\hline\hline
\noalign{\smallskip}
case & $N_\mathrm{d}$ & $N_\mathrm{s}$ & $N_\mathrm{p}$ & $\sigma_*~[\mathrm{km~s}^{-1}]$ & $\eta$\\ 
\noalign{\smallskip}
\hline
\rule[-2mm]{0mm}{0mm}(a)  & 7629 & 13026 & 38785 & $150^{+24}_{-18}$ & $0.22^{+0.09}_{-0.09}$\\
\rule[-2mm]{0mm}{0mm}(b)  & 7629 & 18563 & 50521 & $144^{+18}_{-24}$ & $0.22^{+0.12}_{-0.12}$\\
\rule[-2mm]{0mm}{0mm}(c)  & 7629 & 18563 & 49764 & $156^{+24}_{-30}$ & $0.49^{+0.15}_{-0.21}$\\
\hline
\end{tabular}
\label{tab:results}
\end{center}
\end{table}

\subsection{The role of individual source redshifts}
Compared to case (a), individual redshifts for the source galaxies are
omitted in case (b) and replaced by redshifts drawn from a probability 
distribution. Additionally, objects without COMBO-17 redshifts 
are included in the source sample and increase it by about 40\%. 

Table \ref{tab:results} and Fig.\ \ref{fig:measurements} show no significant
change in the best-fit parameters and the 1-$\sigma$ error bars. Recall that
a difference of $6~\mathrm{km/s}$ in $\sigma_\star$ and 0.03 in $\eta$ is just
the resolution of our grid in parameter space. Figure \ref{fig:measurements}
further shows a widening of the 2- and 3-$\sigma$ likelihood contours together
with an extension towards low $\sigma_\star$ and low $\eta$. The comparison of
cases (a) and (b) shows that the tightest constraints are obtained from using
only sources with measured redshift although this implies that potential
sources are ignored. However, the use of redshift probability distributions
instead of individual redshifts for the sources still leads to almost as good constraints.

The next subsections give some more tests on the influence of source redshifts
on the galaxy-galaxy lensing measurement.

\subsubsection{Source sample from case (a) but with statistical redshifts}
\label{sect:sources_sub1}

The derivation of the redshift probability distribution in Sect.\
\ref{sect:z_estimates} takes the magnitude- and redshift-dependent
completeness of the redshifts in COMBO-17 into account \citep[see][ Fig.\
14]{wolf2004}. For the magnitudes of the lens sample, the completeness reaches
100\% at all redshifts. For the fainter magnitudes of the sources, however,
the completeness is a function of redshift. Here, the completeness is largest
for $z\approx 1$ and declines toward lower and higher redshift. Therefore,
the redshift distribution of all sources with COMBO-17 redshift does not
follow the probability distribution given in Eqs.\
(\ref{eq:Prob_z})--(\ref{eq:zm_mr}) and we can thus not use the same source
sample as in case (a) and assign random redshifts from Eqs.\
(\ref{eq:Prob_z})--(\ref{eq:zm_mr}). Indeed, when trying this we obtained a
best-fit $\sigma_\star$ that was 1-$\sigma$ above the best-fit from case (a).

\subsubsection{Statistical redshifts only for sources without COMBO-17
  redshift}
\label{sect:sources_sub2}
We use the same source selection here as in case (b). The difference is that
here COMBO-17 redshifts are used whenever they are available for a
source. Only those sources without COMBO-17 redshift are assigned a random
redshift. The best-fit parameters remain unchanged compared to case (b) but
the likelihood contours become tighter than in case (b). The upper limits (1-3
$\sigma$) on $\sigma_\star$ remain almost unchanged, but the lower limits
increase. The 3-$\sigma$ lower limit is increased from
$36~\mathrm{km~s}^{-1}$ in case (b) to $60~\mathrm{km~s}^{-1}$ here. On the
other hand, the contours are still slightly wider than in case (a). This leads
to two conclusions: (1) The comparison of cases (a) and (b) shows that
knowledge of individual source redshifts only slightly improves the
constraints by excluding very small values of the velocity dispersion. 
(2) The comparison with case (a) also shows that the inclusion of
additional sources without COMBO-17 redshift does not improve but
rather weakens the constraints. This result appears, at first sight, 
somewhat surprising, yet can probably be explained by the fact that
objects without COMBO-17 redshifts are mainly faint. But faint objects
have less accurate and useful shape measurements than brighter ones
and therefore hardly influence the constraints.

Strictly speaking we should not apply the redshift distribution from Sect.\
\ref{sect:z_estimates} to the sample of sources without COMBO-17 redshifts for
the reasons detailed in Sect.\ \ref{sect:sources_sub1}. However, these
account for only about 30\% of all sources used in this
case. Furthermore, they seem to have too noisy shape measurements, and
thus little weight, so that the slightly wrong redshift
probability distribution does not play a significant role.

\subsubsection{Median instead of random redshifts}
\label{sect:sources_sub3}
Given the small influence individual source redshifts seem to have we
explored how the results would change if we do not use Monte Carlo integration
to deal with the redshift probability distribution. Instead, we assign each
source of case (b) a fixed redshift which is just the median redshift at the
magnitude of the source, see Eq.\ (\ref{eq:zm_mr}). The advantage is that then
the computation is simplified and becomes much faster than requiring
50 Monte Carlo realizations. Compared to case (b) we find a
small shift of the likelihood contours towards higher $\eta$. The shift is 1-2
gridpoints, so $\Delta\eta=0.03-0.06$. In the direction of $\sigma_\star$ we
do not see any change.

\subsubsection{Sensitivity to changes in the redshift probability
  distribution}
\label{sect:sources_sub4}
Instead of the redshift distribution derived from the COMBO-17 data itself we
now test the redshift probability distribution given by
\citet{brainerd1996}. This probability distribution assumes a linear relation
between magnitude and redshift. At $R=24$ it predicts a roughly 20\% smaller
redshift than our model, yielding only a 10\% change in the angular
diameter distance in our cosmology. At fixed lens redshift $z_\mathrm{d}$, the
ratio $D_\mathrm{ds}/D_\mathrm{s}$ also becomes smaller with
decreasing source redshift $z_\mathrm{s}$. Equation (\ref{eq:SIS_gamma})
shows that underestimating the source redshift will overestimate the velocity
dispersion $\sigma_\star$. Indeed we measure, using the same sources as in
case (b), a larger best-fit $\sigma_\star=156~\mathrm{km/s}$ when using the
redshift probability distribution from \citet{brainerd1996}, compared to
$\sigma_\star=144~\mathrm{km/s}$ for case (b). This shift is not significant
given the large error bars of our measurement of galaxy-galaxy
lensing. However, it might be significant for larger surveys with better
statistics like the Sloan Digital Sky Survey \citep[e.g.][]{guzik2002} or the
Red Sequence Cluster Survey \citep[e.g.][]{hoekstra2004}.

\subsection{Importance of individual lens redshifts}
Going from case (b) to (c), individual redshifts for the lenses are omitted as
well and replaced by redshifts drawn from a probability
distribution. Consequently, the luminosity measurement from COMBO-17 is
replaced by luminosities estimated from apparent magnitudes and redshifts as
described in Sect.\ \ref{sect:z_estimates}.

Table \ref{tab:results} and Fig.\ \ref{fig:measurements} show a dramatic
change in the likelihood contours. The best-fit velocity dispersion increases
from $\sigma_\star=144~\mathrm{km~s}^{-1}$ to
$\sigma_\star=156~\mathrm{km~s}^{-1}$ and the 1-$\sigma$ error increases by
about 30\%. The shift towards higher velocity dispersion is not
significant. The best-fit $\eta$, on the other hand, becomes significantly
larger (2-$\sigma$ significance) and the 1- 3$\sigma$ errors increase by about
50-100\%.

As for the sources, we perform some tests on the influence of lens redshifts
on the galaxy-galaxy lensing measurement.

\subsubsection{Influence of the luminosity estimation}
\label{sect:lenses_sub1}
First, we investigate whether there is any difference between using the
luminosities derived from COMBO-17 photometry and the luminosities estimated
using Eq.\ (\ref{eq:L}) with the correct redshift and apparent magnitude of a
lens; this assumption enters only in Eq.\ (\ref{eq:SIS_tf}). Surprisingly, the
likelihood contours remain almost unchanged and also the best-fit values just change by $\Delta\eta=0.03$ towards smaller $\eta$. It seems that the
luminosities of the lenses do not have to be known very accurately.

\subsubsection{Influence of lenses with too low redshift estimate}
\label{sect:lenses_sub2}
When using random redshifts for the lenses in case (c), the redshifts of some
lens candidates will be underestimated so that these galaxies are considered
as lenses although they lie behind the actual source. Here, we exclude all
such pairs, that is, we perform a measurement as in case (c) but use only those
lens-source pairs for which the COMBO-17 redshift of the lenses is lower than
the redshift estimate of the sources. This reduces the number of pairs from 
49764 to 44738 while the number of lenses and sources remains the same as in 
case (c). The best-fit parameters become
$\sigma_\star=156^{+24}_{-24}~\mathrm{km/s}$ and $\eta=0.49^{+0.15}_{-0.24}$.
The contours become slightly tighter than in case (c) but there is no big
change. 

\subsubsection{Influence of non-galaxy lenses}
\label{sect:lenses_sub3}
So far, in all our measurements we used the same set of lenses. In
particular, these are only objects which are classified as galaxies or
likely galaxies. However, in the absence of individual redshifts,
typically no such classification is available. Therefore, we here use
all objects with $R=18-22$ as lenses regardless of their
classification by COMBO-17. Furthermore, we exclude all unresolved
objects from the lens sample. The new lens sample contains 7719
objects, of which 7405 were already used in cases (a)-(c). 224 lenses
from cases (a)-(c) are excluded here because they are unresolved. On
the other hand, 314 lenses are used here but were not considered in
cases (a)-(c). Of these, 217 are classified as galaxies and have
redshift estimates so that they could in principle have been used as
lenses. These 217 galaxies did not enter the lens sample of case (a)
because they have no source from case (a) within $35\arcsec$. The
remaining 97 lenses that are used here but not in cases (a)-(c) are 28
stars, 30 quasars, 4 strange objects and 35 galaxies without redshift
estimates. Therefore, the contamination of our lens sample here from
non-galaxy objects is below 1\%. The likelihood contours shift towards
smaller $\sigma_\star$ and larger $\eta$ compared to case (c) and the
contours become wider. The best-fit parameters become
$\sigma_\star=144^{+24}_{-30}~\mathrm{km/s}$ and
$\eta=0.52^{+0.18}_{-0.21}$. When leaving out the 62 non-galaxy lenses
the likelihood contours remain essentially the same. This shows that
the shift of the likelihood contours compared to case (c) is not due
to the inclusion of non-galaxy objects but to a change in galaxy
population of the lens sample.

\subsubsection{COMBO-17 redshifts with COMBO-17 errors}
\label{sect:lenses_sub4}

To explore how precise individual lens redshift measurements should
be, we use the redshift errors from COMBO-17 to assign random
redshifts with increasing errors to the lenses. Galaxies in the
magnitude range $R=18-22$ have redshift errors $\sigma_z\leq 0.223$
with a mean of about $\langle \sigma_z \rangle
\approx 0.02$. Random redshifts are assigned to each lens with redshift $z$
and redshift error $\sigma_z$ using a gaussian distribution with mean
$z$ and width $f\times \sigma_z$. Even for $f=10$ the likelihood
contours are only slightly wider than those in case (b). It is
therefore apparent that the error on the lens redshifts can become
fairly large without changing the results significantly. However, by
design, each lens gets on average the correct redshift assigned so
that this statement only holds true for unbiased estimates of the
mean.

\subsubsection{Redshifts from $UBVRI$ only}
\label{sect:lenses_sub5}
Finally, we investigate a case intermediate between case (b) where full
redshift information for the lenses is available, and case (c) without any
redshift information. We assign redshifts to the sources as in case (b) while
for the lenses we use redshifts derived from only the 5 broad-band filters
$UBVRI$. The redshifts are estimated using exactly the same technique as for
17 filters. 

First, we briefly summarize the differences between the redshifts and
classification from 5 and 17 filters for potential lenses, i.e.\ objects with
$R=18-22$. From the 10908 objects in that magnitude range 8124 objects are
classified (on the basis of their SED) as galaxies both times. Only 21 objects
are classified as galaxies in the 17-filter classification but missed in the
5-filter classification. From just broad-band filters these 21 objects are
mainly classified as quasars (15 objects). Conversely, 733 objects are
classified as galaxies from the broad-band filters although with 17 filters
they are identified as mainly stars (697 objects) but also quasars (32
objects) or strange objects (4 objects). The redshift difference for objects
classified as galaxies both times is on average only $\langle
z_5-z_{17}\rangle=-0.017\pm 0.074$, where the subscript indicates the number
of filters used for the redshift estimation. The difference in the SDSS
$r$-band restframe magnitude $M_r$ is on average $0.12\pm 0.67~\mathrm{mag}$
implying that on average the luminosities are 10\% smaller when derived from 5
filters. However, the scatter is very large so that the luminosity estimates
can differ by great factors for individual objects. The typical magnitude
difference is $\sqrt{\langle (M_{r,5}-M_{r,17})^2\rangle}=0.68$. This implies
that the luminosities are typically wrong by almost a factor of 2!

\begin{figure}
  \begin{center}
   \leavevmode
      \includegraphics[angle=-90,width=\hsize]{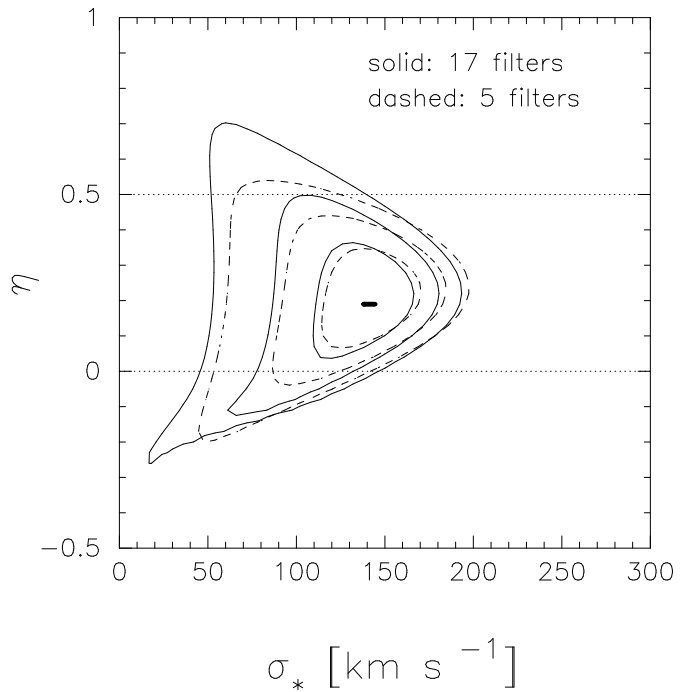}

   \caption{Likelihood contours for the lens model described in Sect.\
     \ref{sect:method}, contours are 1-, 2- and 3-$\sigma$. Solid lines
     correspond to a measurement where all 17 filters are used for redshift
     estimation and classification, dashed contours show the same with
     redshift estimation and classification based on just five broad-band
     filters.}
   \label{fig:photz5_17}
   \end{center}
\end{figure} 

Figure \ref{fig:photz5_17} shows likelihood contours obtained from the
17-filter classification and from the 5-filter classification with the
same set of 7577 lens and 18553 source galaxies being used. In both
cases, the redshifts of the sources are estimated from their
magnitudes as in case (b). The best-fit parameters with 1-$\sigma$
errors are $\sigma_\star=144^{+18}_{-24}~\mathrm{km/s}$ and
$\eta=0.19^{+0.12}_{-0.12}$ for the 17-filter classification and
$\sigma_\star=144^{+24}_{-24}~\mathrm{km/s}$ and
$\eta=0.19^{+0.12}_{-0.09}$ for the 5-filter classification. The
agreement between the two measurements is very good. Surprisingly, the
constraints on $\eta$ are even tighter from the 5-filter
classification than when using the full COMBO-17 filterset. The
differences appear just for large values of $\eta$. However, the upper
bound of $\eta$ is determined from the brightest lenses which would
be weighted highly (from Eq.\ (\ref{eq:SIS_tf})) for large
$\eta$. Therefore, excluding the brightest lenses does increase the
upper bound on $\eta$. Although, on average, luminosities are
underestimated in the 5-filter classification, we see exactly the
opposite for bright objects. For example, 235 objects have
$L>5L_\star$ in the 5-filter classification. On average, these 235
objects are 0.29 mag brighter than in the 17-filter classification so
that their luminosities are overestimated by about 30\%. For the 60
objects with $L>10L_\star$ this difference becomes 1.16 mag or a
factor of almost 3 in luminosity. Therefore, we think that the tighter
contours derived from the 5-filter classification do not mean that
better contraints can be obtained but rather shows the limitations of
this smaller filter set compared to the whole 17-filter classification.

\section{Summary and outlook}
\label{sect:outlook}

We used the COMBO-17 survey to investigate the importance of individual
redshift measurements on galaxy-galaxy lensing studies. We find that redshift
information for the lens galaxies plays a crucial role in constraining their
dark matter halos while for the source galaxies it is sufficient to have
roughly correct redshift probability distributions available. We have also
seen that redshifts obtained from just $UBVRI$ ($\langle
\sigma_z/(1+z)\rangle =0.03$ for $R=18-22$) instead of the full 17 filter set
available in COMBO-17 ($\langle \sigma_z/(1+z)\rangle =0.015$ for
$R=18-22$) are sufficient to constrain the redshifts of the lenses and their
rest-frame luminosities.  

However, some benefits from the 17-filter classification remain. Most
importantly, the 17-filter classification was used to derive the
redshift probability distribution of the source population. In deep
weak lensing studies, sources are typically fainter than the magnitude
limit of current galaxy redshift surveys. Therefore, any redshift
probability distribution can only be obtained from extrapolation to
faint magnitudes. In Sect.\ \ref{sect:sources_sub4} we have shown 
that this extrapolation can introduce a bias into the measurement that
is of the same order or even larger than the errors from large surveys
like the RCS or SDSS. In principle, the redshift distribution of the
sources could also be derived from just 5 filters. However, from 5
filters, the decrease in redshift accuracy is larger for the fainter
sources than for the brighter lenses. Furthermore, fewer sources get
reliable redshift estimates from 5 than from 17 filters. Therefore,
the source redshift distribution derived from 5 filters will be less
accurate than from 17 filters.

The gain from multi-colour or spectroscopic data for the lenses will
even be larger than shown from the comparisons in this
paper. Knowledge of individual redshifts will allow one to select
lenses and sources not from less reliable magnitude cuts but from the
redshifts themselves. Redshifts will also allow one to include fainter
lenses or, when redshifts are available for the sources as well,
brighter sources and will thus improve the statistics.

Any information on the nature of the lens galaxies will additionally
allow one to investigate if and how the properties of the dark matter
halos depend on galaxy classes which can be defined according to e.g.\
luminosity, colour, stellar mass or environment. Such investigations
are necessary to test and improve our understanding of galaxy
formation and evolution where dark matter is supposed to play a major
role.

\begin{acknowledgements}
CW was supported by a PPARC Advanced Fellowship. MK acknowledges support by
the BMBF/DLR (project 50 OR 0106), by the DFG under the project SCHN 342/3--1,
and by the DFG-SFB 439

\end{acknowledgements}

\bibliographystyle{aa}
\bibliography{bib}

\end{document}